\def\EPJ{{\em Eur. Phys. J.} C}
\documentclass{ws-p8-50x6-00}

\def\ifmath#1{\relax\ifmmode #1\else $#1$\fi}%
\def\rb{\ifmath{{\mathrm{b}}}}
\def\rB{\ifmath{{\mathrm{B}}}}
\def\rc{\ifmath{{\mathrm{c}}}}

\def\rD{\ifmath{{\mathrm{D}}}}
\def\rK{\ifmath{{\mathrm{K}}}}
\def\rL{\ifmath{{\mathrm{L}}}}

\def\rp{\ifmath{{\mathrm{p}}}}
\def\rq{\ifmath{{\mathrm{q}}}}
\def\rV{\ifmath{{\mathrm{V}}}}
\def\rZ{\ifmath{{\mathrm{Z}}}}

\def\syst.{\ifmath{{\mathrm{syst.}}}}
\def\stat.{\ifmath{{\mathrm{stat.}}}}

\begin{document}

\title{Polarization and spin alignment in multihadronic Z$^0$ decays}

\author{T.H.~Kress}

\address{Department of Physics, University of California, Riverside,
USA\\OPAL Collaboration, E-mail: Thomas.Kress@cern.ch}

\maketitle

\abstracts{
The large statistics of millions of hadronic $\rZ^{0}$ decays, accumulated 
by the four LEP experiments between 1989 and 1995, allowed for detailed
investigations of the fragmentation process.
Inclusive $\Lambda_{\rb}$ baryons and $\Lambda$ hyperons at intermediate 
and high momentum have been found to show longitudinal polarization.
This may be related to
the primary quark and antiquark polarization and the hadronization
mechanism which produces the leading baryons.
Helicity density-matrix elements have been measured for a variety of
vector mesons produced inclusively in hadronic $\rZ^{0}$ decays. The 
diagonal elements of some of the light mesons and the $\rD^{*\pm}$ show a
preference for a helicity-zero state if the meson carries a large 
fraction of the available energy. The mechanism which produces such spin
alignment in the non-perturbative hadronization of the primary partons
to the vector mesons is so far unexplained. For the $\rB^{*}$, the results
are consistent with no spin alignment, which is expected in a picture
based on HQET. For some meson species non-diagonal elements have 
been measured indicating coherence phenomena due to final-state 
interaction between the primary quark and antiquark.
}
\section{Introduction}
Between 1989 and 1995, the four LEP collaborations have recorded over 16 
million hadronic $\rZ^{0}$ decays which allow for detailed studies of the
hadronization process.

Primary quarks produced in e$^{+}$e$^{-}$ annihilations around the $\rZ^{0}$ 
pole are strongly left-handed with longitudinal polarizations of
approximately $-0.91$ for down-type quarks and $-0.64$ for up-type quarks.

A measurement of the leading baryon polarization in self-analysing
weak decay modes can be used to investigate the polarization 
transport in the fragmentation process and gives useful new information
on the dynamics of the hadronization process. 

The vector mesons are $\rq\bar{\rq}$ systems of total spin one without
angular momentum, and any alignment of the meson spin must arise at
least in part from the dynamics of the hadronization phase. 

\section{Principles of the Measurements}
To measure the polarization of $\Lambda$ baryons, the decay angular 
distribution of the proton in the hyperon rest frame in the decay
$\Lambda \rightarrow \rp\pi$ is analysed.\cite{bib:pol_l_a,bib:pol_l_o}

The $\Lambda_{b}$ polarization is studied using its semileptonic
decays with a $\Lambda$ hyperon reconstructed in the final 
state.\cite{bib:pol_lb_a,bib:pol_lb_d,bib:pol_lb_o}
The energy spectra of the charged lepton and of the neutrino, 
determined by the missing energy in the hemisphere containing the 
$\Lambda\hspace{0.25mm}l$ system, are sensitive to the polarization.

Spin alignment of vector mesons can be described in terms of the spin
density matrix $\rho_{\lambda\lambda^{\prime}}$, which is a 3x3 Hermitian 
matrix with unit trace, usually defined in the helicity base.
As described in~\cite{bib:sa_o1} the elements of the helicity density 
matrix can be measured using the angular distribution of the corresponding
vector meson decay products.
The diagonal elements, limited by $0\le\rho_{\lambda\lambda}\le 1$,
represent the relative intensities of the three helicity states 
$\lambda = -1,0,+1$. 
In parity conserving decays, only one diagonal element is measurable. 
Thus, a state with no spin alignment corresponds to $\rho_{00} = 1/3$. 
The off-diagonal elements, for example $\rho_{1-1}$, measure the 
correlation between states with different helicities. 

\section{Polarization of Baryons}
\subsection{Polarization of $\Lambda$ hyperons}
\begin{figure}[h]
\epsfxsize=15.0pc \epsfysize=14.6pc 
\epsfbox{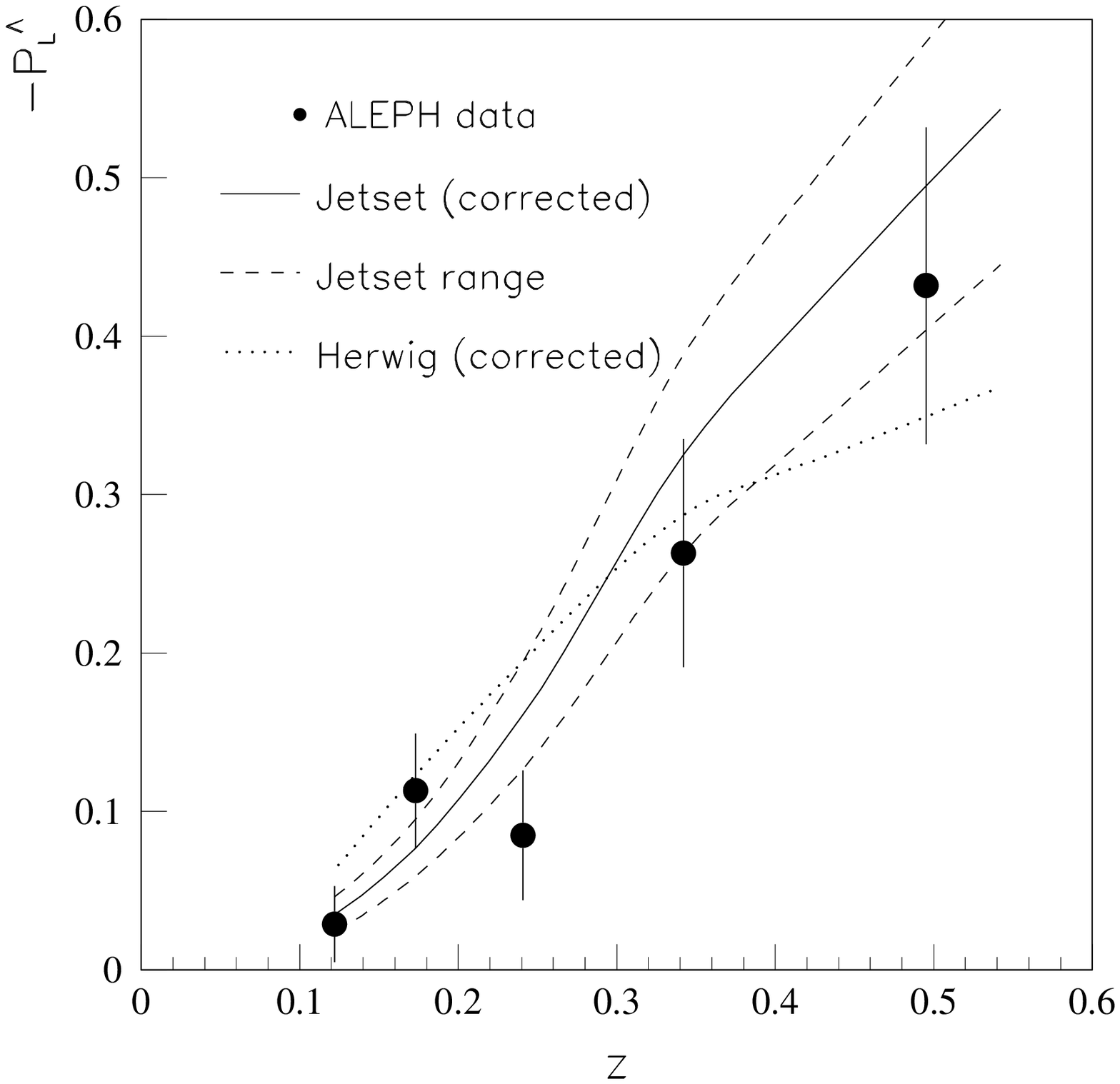}
\epsfxsize=14.0pc \epsfysize=14.5pc 
\epsfbox{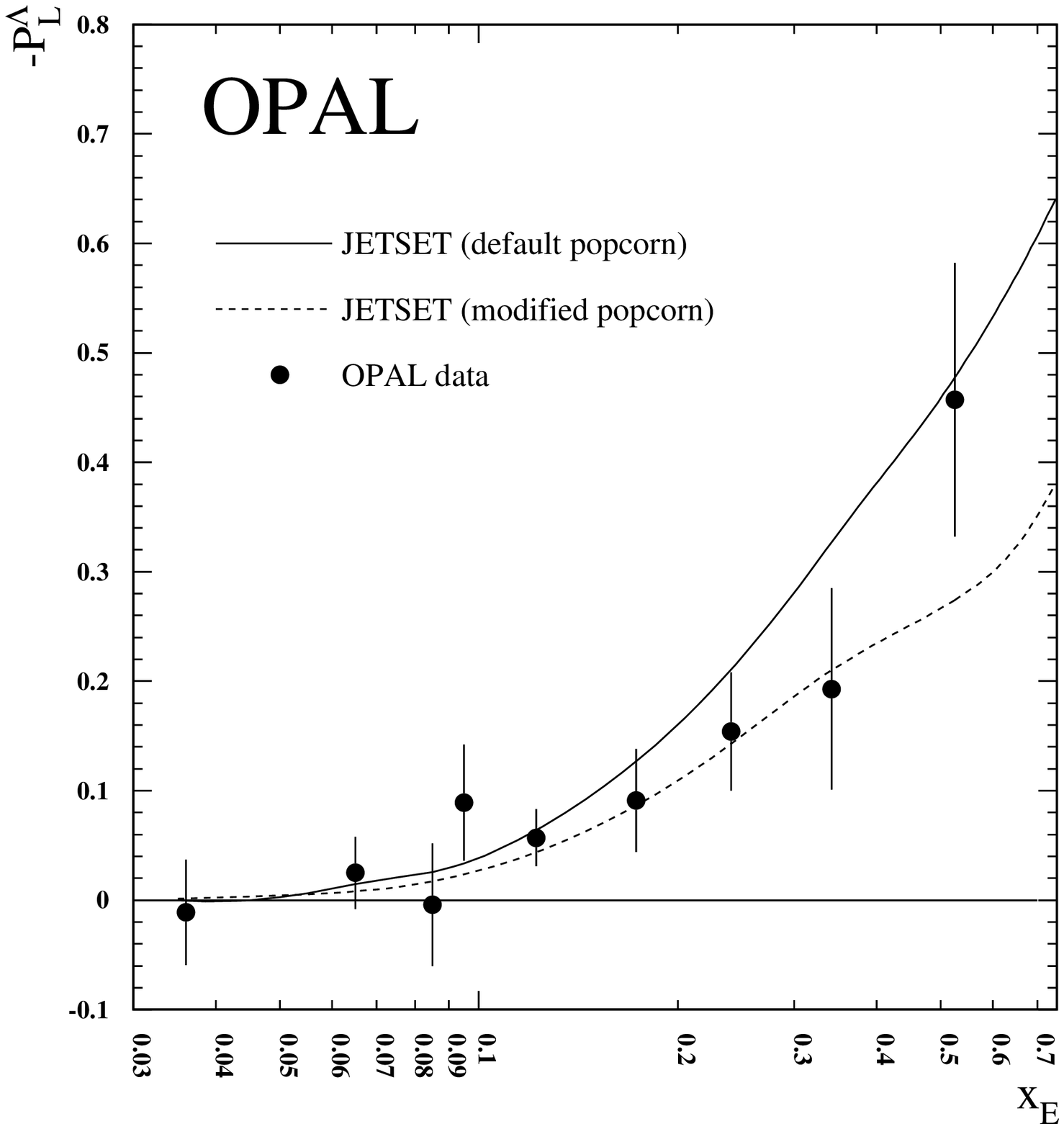}
\caption{Measured longitudinal polarization of $\Lambda$ hyperons
as a function of the scaled energy. The curves show the prediction 
of a model discussed in the text. \label{fig:pol_lambda}}
\end{figure}

Figure \ref{fig:pol_lambda} shows the ALEPH\cite{bib:pol_l_a} and 
OPAL\cite{bib:pol_l_o} measurements of the $\Lambda$ polarization for
different 
ranges of the $\Lambda$ energy scaled to the beam energy (z$\equiv$$x_{E}$).
For $x_{E} > 0.3$, the ALEPH collaboration measures a polarization of 
$-0.32 \pm 0.07$
consistent with the corresponding OPAL result of $P_{\rL} = -0.329 \pm 0.076$.

A model of Gustafson and H\"akkinen\cite{bib:pol_l_gh} is used to calculate 
the expected $x_{E}$ dependence of the polarization using the JETSET and HERWIG
Monte Carlo programs to determine the $\Lambda$ production rates from 
several different sources. In this model the following assumptions were made:
\begin{itemize} 
\item the spin of the $\Lambda$ is determined by the spin of
      the s quark and directly produced $\Lambda$ are polarized
      as the primary s quark;
\item $\Lambda$ particles which are decay products of heavier baryons
      containing the primary s quark will have a fraction of the
      polarization as given by the SU(6) model; 
\item any polarization is lost if a primary u or d quark becomes a 
      constituent quark of the $\Lambda$, forming a spin-0 ud diquark;
\item $\Lambda$ hyperons containing only quarks produced in the
      fragmentation are not polarized.
\end{itemize}
This relatively simple constituent quark model fits the observations.
This supports the idea that the primary strange quark polarization is
transferred completely to the formed final state hyperon during 
hadronization.

However, other interpretations\cite{bib:pol_l_kbh} of the $\Lambda$ 
polarization are by no means disfavoured.
 
\subsection{Polarization of $\Lambda_{\rb}$ baryons}
$\Lambda_{\rb}$ polarization has been studied by ALEPH\cite{bib:pol_lb_a},
DELPHI\cite{bib:pol_lb_d} and OPAL.\cite{bib:pol_lb_o}
\begin{table}[h]
\caption{ $\Lambda_{\rb}$ polarization measurements. \label{tab:a}}
\begin{center}
\begin{tabular}{|l|l|}
\hline
$\rule[-2mm]{0mm}{6mm}
-0.23\hspace{0.5mm}^{+0.24}_{-0.20}(\stat.)
 \hspace{0.95mm}^{+0.08}_{-0.07}\hspace{2.15mm}(\syst.)$ & ALEPH \\
$\rule[-2mm]{0mm}{6mm}
-0.49\hspace{0.5mm}^{+0.32}_{-0.30}(\stat.)\pm 0.17(\syst.)$ & DELPHI \\
$\rule[-2mm]{0mm}{6mm}
-0.56\hspace{0.5mm}^{+0.20}_{-0.13}(\stat.)\pm 0.09(\syst.)$ & OPAL \\
\hline
\end{tabular}
\end{center}
\end{table}

A model of Falk and Peskin,\cite{bib:pol_sa_fp} based on HQET, taking 
account of b baryons proceeding through intermediate states such as 
$\Sigma_{\rb}$ and $\Sigma_{\rb}^{*}$ gives a range of expected polarization
between $-0.54$ and $-0.88$.\cite{bib:pol_lb_o}

The experimental results given in Table~\ref{tab:a} agree with each other 
and with the hypothesis of the HQET model that there is no loss of the 
initial b quark polarization during the fragmentation.

\section{Vector Meson Helicity Density Matrix Elements}
Helicity density-matrix elements have been measured by ALEPH,\cite{bib:sa_a}
DELPHI\cite{bib:sa_d1,bib:sa_d2} and
OPAL\cite{bib:sa_o1,bib:sa_o2,bib:sa_o3} for various meson species.
\begin{figure}[h]
\begin{center}
\epsfxsize=24.0pc 
\epsfbox{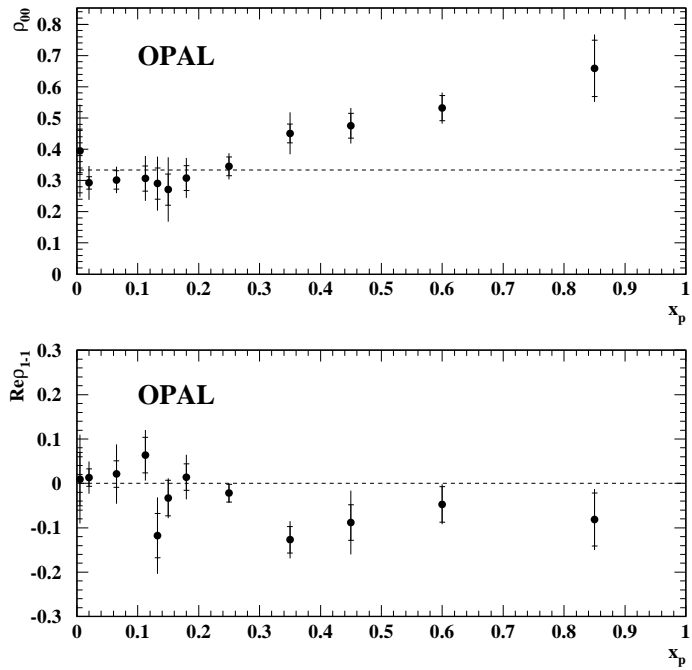}
\caption{$\rK^{*0}$ helicity density matrix elements, $\rho_{00}$ 
and Re($\rho_{1-1}$) as functions of $x_{p}$. \label{fig:sa_kstar_opal}}
\end{center}
\end{figure}
As an example, the results of the inclusive $\rK^{*}(892)^{0}$ 
analysis\cite{bib:sa_o2} are shown in Fig.~\ref{fig:sa_kstar_opal}. 
At large momentum where the $\rK^{*}$ most likely contains a primary quark 
from the $\rZ^{0}$ decay, the results show a clear deviation from 
$\rho_{00} = 1/3$ and Re($\rho_{1-1}$) = 0. 

\begin{figure}[h]
\begin{center}
\epsfxsize=24.0pc 
\epsfbox{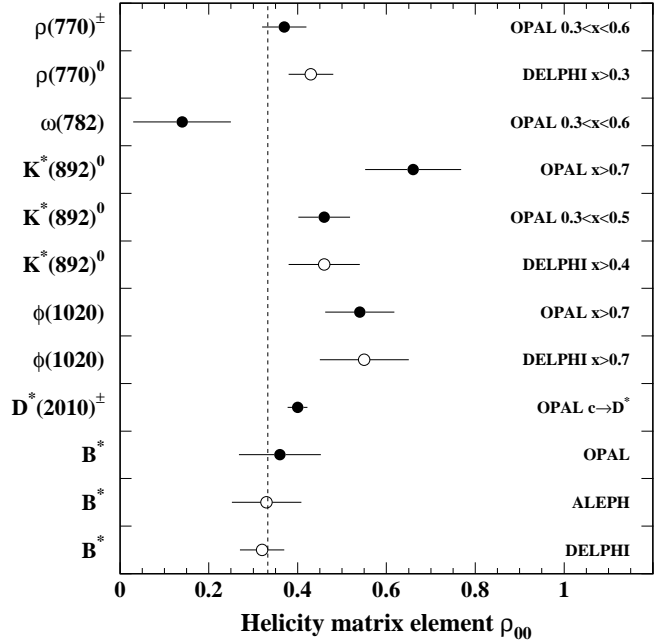}
\caption{Values of helicity density matrix element $\rho_{00}$ for various
meson species. \label{fig:sa_summary}}
\end{center}
\end{figure}

Figure \ref{fig:sa_summary} summarizes the measurements of the matrix
element $\rho_{00}$. 
$\rB^{*}$ mesons show no deviation from equal population of the three helicity
states. 
The $\rD^{*}(2010)^{\pm}$ sample consists of both directly produced 
$\rc \rightarrow \rD^{*}$ mesons and those from excited charm hadron decays.
A value of $\rho_{00} = 0.40 \pm 0.02$ indicates an enhanced production in
the zero helicity state. 
The results for the lighter leading mesons, which are likely to contain
the primary polarized quark, are not completely conclusive. However, in 
general there is a clear indication for a production in the helicity zero
state.

A statistical model based on spin counting\cite{bib:pol_sa_fp}
fits the $\rB^{*}$ measurements. 
A QCD-inspired model\cite{bib:sa_ar} predicts $\rho_{00} = 0$ for leading
vector mesons in the limit that quark and meson masses and transverse
momentum can be neglected.
If the vector meson production is considered as arising from a 
helicity-conserving process $\rq \rightarrow \rq\rV$, then $\rho_{00} = 1$ is 
predicted.\cite{bib:sa_d}
In the generally successful JETSET string model of hadron formation, no
spin alignment is expected and the same is true for the HERWIG cluster
model. 
The pattern of observations would benefit from more experimental
measurements, and awaits a firm theoretical interpretation.

For some mesons OPAL\cite{bib:sa_o1,bib:sa_o2} and DELPHI\cite{bib:sa_d2} 
have measured non-diagonal helicity density matrix elements. 
Whilst the OPAL collaboration has some evidence for small negative values
of Re($\rho_{1-1}$) for the $\rD^{*}$, $\phi(1020)$ and $\rK^{*}(892)^{0}$ 
mesons, which is in agreement with some theoretical 
expectations\cite{bib:sa_abmq} that coherence phenomena due to final-state
interaction between the primary quark and antiquark play a role in 
fragmentation, 
the DELPHI results for $\rho^{0}$, $\rK^{*}(892)^{0}$ and $\phi(1020)$ are
consistent with zero.

\section*{Acknowledgments}
I would like to thank the organizers of the 30th International Symposium on
Multiparticle Dynamics for the invitation to present these LEP results
during the very pleasant meeting in Tihany, Hungary.


\end{document}
